\documentclass[a4paper,onecolumn,11pt]{quantumarticle}
\pdfoutput=1
\usepackage[utf8]{inputenc}
\usepackage[english]{babel}
\usepackage[T1]{fontenc}
\usepackage{amsmath}

\usepackage{tikz}
\usepackage{lipsum}
\usepackage[numbers]{natbib}
\usepackage{chemformula} % Formula subscripts using \ch{}
\usepackage[T1]{fontenc} % Use modern font encodings
\usepackage[utf8]{inputenc}
\usepackage[english]{babel}
\usepackage{amsmath}
\usepackage[caption = false]{subfig}
\usepackage{graphicx,epstopdf}
\usepackage{blindtext}
\usepackage{lipsum}
\usepackage{amsfonts}
\usepackage{bbm}
\usepackage{amssymb}
\usepackage{enumerate}
\usepackage{color}
\usepackage{latexsym}
\usepackage{times,txfonts}
\usepackage{wrapfig}
\usepackage{ulem}
\usepackage{orcidlink}
\usepackage{caption}
\usepackage{subfig}
\usepackage{cancel,soul}
\usepackage{soul}
\usepackage{xcolor}
\newcommand{\redsout}[1]{\textcolor{red}{\st{#1}}}
\newcommand{\ave}[1]{\langle#1\rangle}
\newcommand{\ket}[1]{|#1\rangle}
\newcommand{\bra}[1]{\langle #1|}

\begin{document}

\title{Quantum Imaging and Metrology with Undetected squeezed Photons: Noise Canceling and Noise Based Imaging}

\author[1]{Saheb Samimi}
\author[1]{Zahra Ghasemi}
\author[1,2]{Hamidreza Mohammadi}
\email{hr.mohammadi@sci.ui.ac.ir}
\affiliation[1]
{Center of Quantum Science and Technology (CQST), University of Isfahan, Isfahan}
\affiliation[2]
{Faculty of Physics, University of Isfahan, Isfahan}
%\author{Lídia del Rio}
%\affiliation{Institute for Theoretical Physics, ETH Zurich, 8093 Zurich, Switzerland}
%\orcid{0000-0002-2445-2701}
%\author{Christian Gogolin}
%\email{latex@quantum-journal.org}
%\homepage{http://quantum-journal.org}
%\orcid{0000-0003-0290-4698}
%\thanks{You can use the \texttt{\textbackslash{}email}, \texttt{\textbackslash{}homepage}, and \texttt{\textbackslash{}thanks} commands to add additional information for the preceding \texttt{\textbackslash{}author}. If applicable, this can also be used to indicate that a work has previously been published in conference proceedings.}
%\affiliation{Covestro Deutschland AG, Kaiser-Wilhelm-Allee 60, 51373 Leverkusen, Germany}
%\author{Marcus Huber}
%\affiliation{Institute for Quantum Optics \& Quantum Information (IQOQI), Austrian Academy of Sciences, Boltzmanngasse 3, Vienna A-1090, Austria}
%\orcid{0000-0003-1985-4623}
%\author{Cassandra Granade}
%\affiliation{Microsoft Research, Quantum Architectures and Computation Group, Redmond, WA 98052, USA}
%\author{Johannes Jakob Meyer}
%\affiliation{Dahlem Center for Complex Quantum Systems, Freie Universität Berlin, 14195 Berlin, Germany}
%\orcid{0000-0003-1533-8015}
%\author{Victor V. Albert}
%\affiliation{Institute for Quantum Information and Matter \& Walter Burke Institute for Theoretical Physics, Caltech, Pasadena, CA 91125, USA}
%\orcid{0000-0002-0335-9508}
\maketitle

\begin{abstract}
   In this work a quantum imaging setup based on undetected squeezed photons is employed for metrological applications such as sensitive phase measurement and quantum imaging. Unlike traditional quantum imaging with undetected photons, as introduced by A. Zeilinger \textit{et al.}, the proposed setup incorporates a homodyne detection system and enhances the brightness of the quantum light using optical parametric oscillators (OPOs). The inclusion of OPOs may challenge the validity of the low-gain approximation, necessitating the development of a new theoretical framework that extends beyond this approximation. The results demonstrate a higher signal-to-noise ratio, which serves as a key metric for both image quality and phase-measurement accuracy. Furthermore, an imaging protocol is introduced to effectively suppress background noise. Notably, this protocol enables the extraction of image information encoded in quantum fluctuations (noise), paving the way for non-disruptive imaging. This is particularly significant in the field of bio-imaging, where imaging sensitive living cells with a low damage threshold is of critical importance.
\end{abstract}
Interferometry provides Interferometry serves as a powerful technique to measure for measuring physical quantities with exceptional precision and sensitivity\cite{Frascella:19}.  The interferometry is used in direct detection of gravitational waves \cite{Abbott}. The sensitivity of the  conventional interferometers is bounded by Shot Noise Limit (SNL) \cite{PhysRevA.94.023834,PhysRevD.23.1693,PhysRevA.96.052118}. One of the primary goals in interferometry is to overcome SNL. Using non-classical properties of lights provides promising resources to achieve this  \cite{carranza2012photon,dowling2008quantum,seshadreesan2011parity,lee2012nonlinear,PhysRevA.98.043856,jing2011realization,manceau2017improving,hu2014quantum,bradshaw2018ultimate,zhang2024entanglement,PhysRevA.104.043707}. Recent advancements in nonlinear optics have spurred interest in nonlinear interferometry to achieve highly sensitive and accurate interferometers. \cite{PhysRevLett.128.033602,PhysRevLett.119.223604,xin2021phase,santandrea2023lossy}. In particular, SU(1,1) interferometers, which are constructed by replacing the beam splitters with two mode squeezer devices (e.g  Optical Parametric Amplifiers (OPAs)), offer sensitivity enhancement beyond SNL \cite{chang2020enhanced,PhysRevLett.128.033602,PhysRevLett.118.150401,liu2018loss,tang2024improvement,li2014phase,xu2023phase,lindner2023high,ou2020quantum}. 
Non-degenerate SU(1,1) interferometers operate at distinct wavelengths for interaction and detection parties (e.g. mid-infrared for interaction party and optical spectrum for detection) \cite{jin2024quantum,PhysRevA.101.053843,roeder2024measurement,PhysRevApplied.19.054019}. More precisely, these devices generate two-mode squeezed light at different wavelengths (sinal and idler beams) which enables quantum metrological applications in spectroscopy \cite{kalashnikov2016infrared,lee2019interferometric}, imaging \cite{kviatkovsky2020microscopy,Pearce:23}, optical coherent tomography(OCT)\cite{PhysRevA.94.023834,hudelist2014quantum} and bio-imaging of living cells \cite{buzas2020biological,yang2023interaction}. 

\textit{Quantum Imaging with Undetected Photon} (QUIP), introduced by Zeilinger \textit{et al.}, utilizes an SU(1,1) interferometer. In this scheme, the idler beam interacts with an object, while the image is formed by detecting the signal beams, which never interacts with the object\cite{lemos2014quantum,roeder2024measurement}. In the original setup, a signal-idler photon pair is generated via spontaneous parametric down-conversion (SPDC) in a nonlinear crystal. The idler beam interacts with the object and is subsequently combined with a second idler beam produced in an identical nonlinear crystal. This combination induces interference between the two corresponding signal beams emitted from the respective crystals. Crucially, the observed interference patterns in the signal beams stem from the path indistinguishably (or identity) of the idler beams, rather than any mechanism related to induced emission \cite{PhysRevA.44.4614}.
Theoretical analyses of QIUP typically assume operation in the low-gain regime \cite{PhysRevA.92.013832}. However, this approximation restricts the framework applicability to high-gain systems. In this work, we develop a generalized theoretical model for QIUP that transcends the low-gain assumption. To achieve this, we solve the system dynamics in the Heisenberg picture (as opposed to the Schrodinger picture) and derive analytical expressions for the output creation and annihilation operators at the device’s output port, explicitly incorporating the classical pump field. To improve imaging quality, quantified by the Signal-to-Noise Ratio (SNR), and to investigate the performance of QIUP in high-gain regimes, each nonlinear crystal is placed within a con-focal cavity. This configuration replaces the crystals with two identical Non-degenerate Optical Parametric Oscillators (NOPOs). In each NOPO generate a signal-idler biphoton state via cavity-PDC process\cite{OptCom575131284}, which exhibits significant squeezing, enabling SNR enhancement by leveraging squeezing as a quantum resource. Consequently, we utilize homodyne detection—rather than EMCCD—for data acquisition. We propose an imaging protocol designed to suppress noise. Additionally, we introduce a novel method for constructing images from objects with low damage thresholds, termed Quadrature-Noise Shadow Imaging (QSI). This scheme relies on quadrature quantum noise measurements \cite{cuozzo2022low}.
QSI permits imaging with low photon flux while eliminating the detrimental dark-count noise of sensors in practical setups \cite{cuozzo2022low, barge2022weak,clark2012imaging}. Intriguingly, we demonstrate that image information is encoded into the noise itself, enabling reconstruction via QSI.

This paper is organized as follow: In the next section the setup under consideration is described. Section \ref{sec.3} is devoted to the theoretical description of the model. In first subsection of Sec.\ref{sec.4} opaque objects are characterized while in second subsection transparent objects is addressed. Accordingly two imaging protocols are introduced in Sec. \ref{sec.5}. Finally the conclusion and remarks of this work are provided in Sec. \ref{sec.6}.
\section{The Setup}\label{sec.2}
This section  demonstrates the proposed setup in detail. The schematic diagram of the setup is depicted in figure 1. The imaging process in this figure can be explained through four stages. In the first stage the two mode squeezed state beams, including idler (red) and signal(green) beams, are generated in the output of the first NOPO. Then a dichroic mirror $DM1$ separates them into different paths: The signal (green) beam arrives to the second  NOPO and the idler (red) beam is reflected through $DM1$ toward the object. The idler interacts with the object which is modeled by a lossless beam splitter with transmission $Te^{i\phi_{T}}$ (reflection $Re^{i\phi_{R}}$). In order to provide path identity, a portion of the idler, which is transmitted through the object, is aligned with second idler beam, generated in the second NOPO. Finally, the information about the object is extracted from signal beam due to the quantum interference of idler states. Since the object information are encoded in the signal quadratures, a balanced homodyne detection is applied to the outcome signal of the second NOPO to extract the image information. It is worth noting that undetected light refers to the idler beam as no measurement applied to it. 

\begin{figure}[h!]. 
	\label{fig1}
	\includegraphics[scale=0.35]{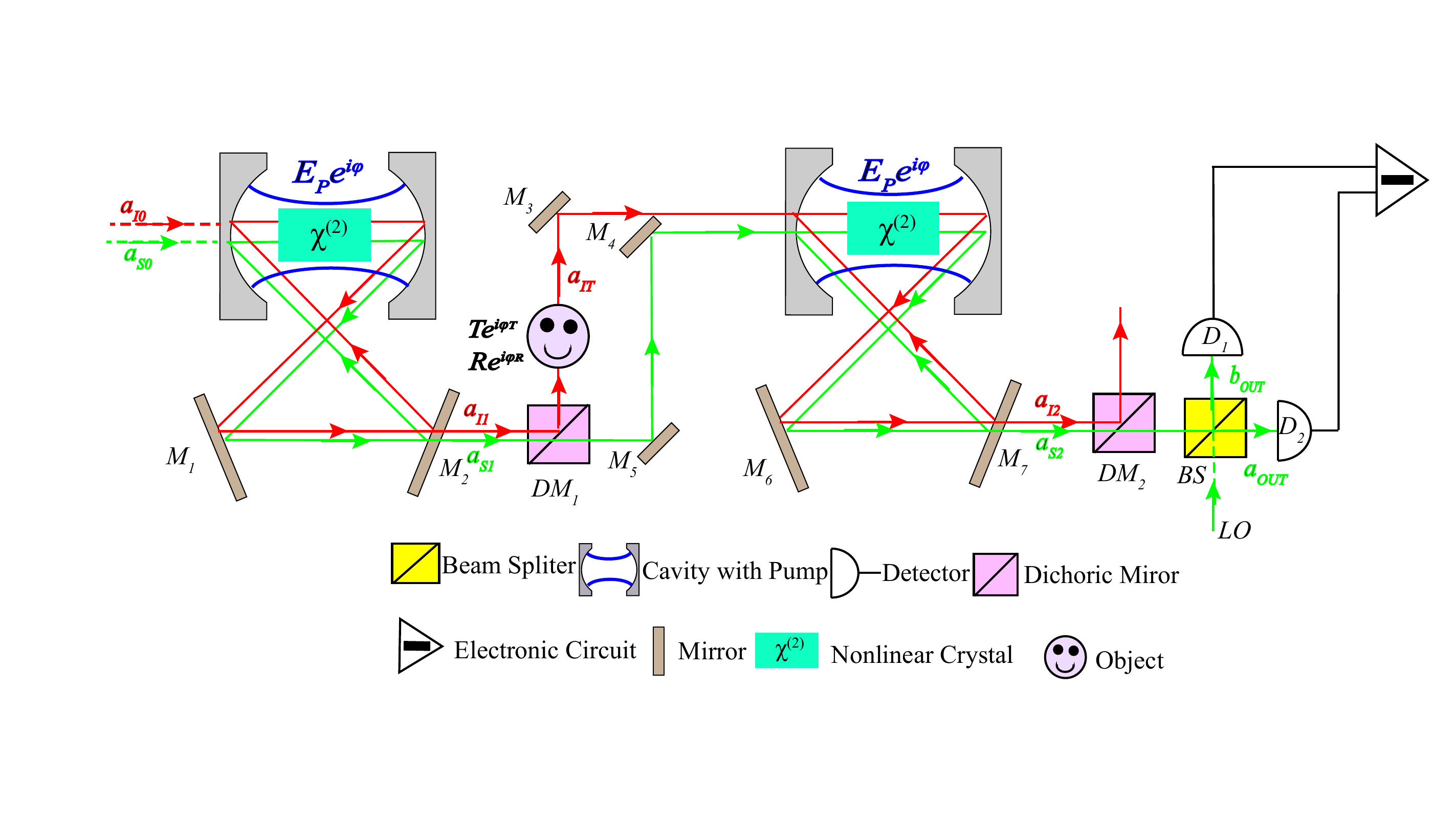}
	\caption{Schematic diagram of the setup}
\end{figure}

In the next section, we present a theoretical framework describing how image information is extracted.
\section{Theoretical Description of The Model}\label{sec.3}
Considering a local oscillator prepared in coherent state, $\ket{\beta e^{i\phi_{\beta}}}$, the output electrical signal of the balanced homodyne detection (depicted in Fig. 1) can be calculated as:
\begin{align}
	\label{eq1}
	S=\frac{1}{4}\beta\left(a_{S2}^{\dagger}e^{-i\phi_{\beta}}+a_{S2}e^{i\phi_{\beta}}\right),
\end{align}
The annihilation (creation) operator $a_{S2}$ ($a_{S2}^{\dagger}$) is correspond to the output signal beam from second NOPO.                                                                                         The image information is encoded in this beam, Assuming that the NOPOs are pumped by classical fields with fixed phase, $\phi_{P1}$ and $\phi_{P2}$, respectively, we can write:
\begin{subequations}
	\label{eq2}
	\begin{align}
		a_{S1}=G_{1} a_{S0}+g_{1} e^{i\phi_{P1}} a_{I0}^{\dagger},\label{eq2a}\\
		a_{I1}=G_{1} a_{I0}+g_{1} e^{i\phi_{P1}} a_{S0}^{\dagger},\label{eq2b}
	\end{align}
\end{subequations}
where the input signal and idler annihilation (creation) operators are denote by $a_{S0}$ ($a_{S0}^{\dagger}$) and $a_{I0}$ ($a_{I0}^{\dagger}$), respectively. In Eqs. \eqref{eq2} the squeezing process is characterized by $G_{1}$ and $g_{1}$. As mentioned above, the object is characterized as a lossless beam splitter. Therefore the transmitted idler field is given by,
\begin{align}
	\label{eq3}
	a_{IT}=Te^{i\phi_{T}}a_{I1}+Re^{i\phi_{R}}a_{IL},
\end{align}   
here, $a_{IL}$ is relate to the empty port of the beam splitter (object). It is straightforward to show that
\begin{align}
	\label{eq4}
	a_{IT}=Te^{i\phi_{T}}\left(G_{1} a_{I0}+g_{1} e^{i\phi_{P1}} a_{S0}^{\dagger}\right)+Re^{i\phi_{R}}a_{IL}.
\end{align}
Two beams ($a_{S1}$ and $a_{IT}$) are then mixed in the second NOPO to built up beam $S_{2}$:
\begin{align}
	\label{eq5}
	a_{S2}=G_{2}a_{S1}+g_{2}e^{i\phi_{P2}}a_{IT}^{\dagger},
\end{align}
where $G_{2}$, $g_{2}$  and $\phi_{P2}$ are the gain parameters and pump phase of the second crystal. Substituting Eqs. \eqref{eq2a} and \eqref{eq4} into Eq. \eqref{eq5} leads to,
\begin{align}
	\label{eq6}
	a_{S2}=Ga_{S0}+ga_{I0}^{\dagger}+ra_{IL}^{\dagger},
\end{align}
with $G=G_{1} G_{2}+g_{1} g_{2}T\exp\left(i\phi_{1}\right)$, $g=g_{1}G_{2}\exp(i\phi_{P1})+G_{1}g_{2}T\exp\left(i\phi_{2}\right)$ and $r=g_{2}R\exp(i\phi_{3})$. Here, the parameters $\phi_{1}=\phi_{P2}-\phi_{P1}-\phi_{T}$, $\phi_{2}=\phi_{P2}-\phi_{T}$ and $\phi_{3}=\phi_{P2}-\phi_{R}$ are defined for simplicity. It is worth noting, since $|G|^{2}-|g|^{2}-|r|^{2}=1$ is satisfied, the commutation relation $\left[a_{S2},a_{S2}^{\dagger}\right]=1$ is also preserved.

\section{System Characteristics}\label{sec.4}
\begin{figure}[h!]
	\centering 
	\subfloat[]{\includegraphics[scale=0.5]{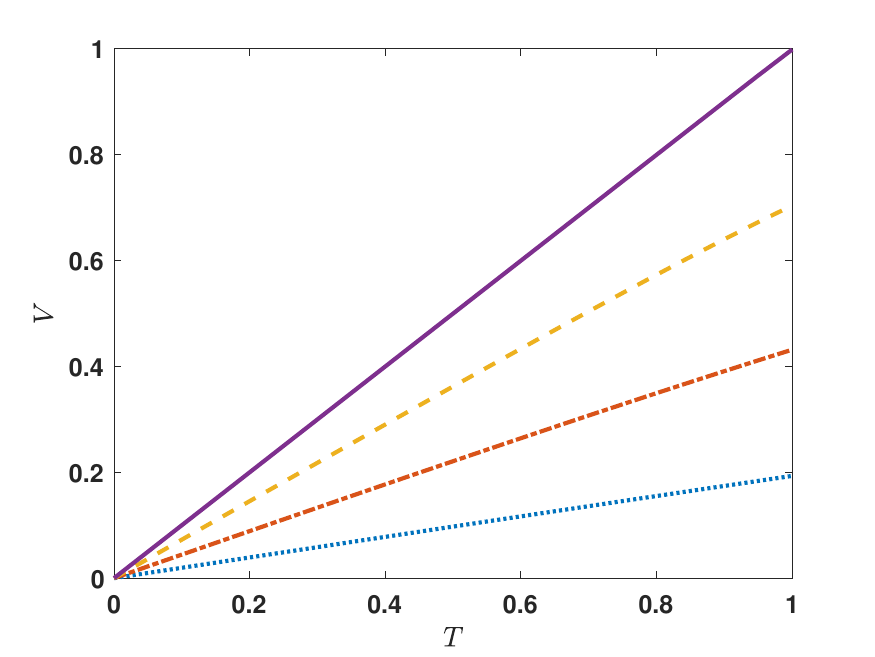}} 
	\subfloat[]{\includegraphics[scale=0.5]{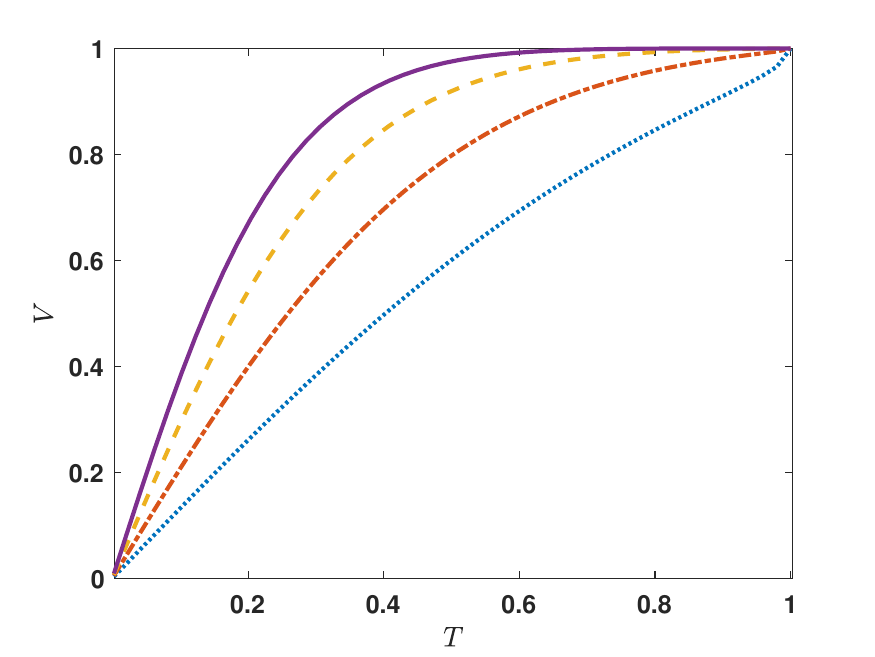}} 
	\caption{Visibility of the image, $\frac{g_1g_2}{G_1G_2}=0.25$ (dot curve), $\frac{g_1g_2}{G_1G_2}=0.5$ (dot-dashed curve), $\frac{g_1g_2}{G_1G_2}=0.75$ (dashed curve) and $\frac{g_1g_2}{G_1G_2}=1$ (solid curve) (a) homodyne detection which is defined by $V=\frac{\max(S^{rms})-\min(S^{rms})}{\max(S^{rms})+\min(S^{rms})}$, where $S^{rms}$ denotes root mean square of output signal. (b) Conventional photon counting which is defined by $V=\frac{\max(n_{S2}^{rms})-\min(n_{S2}^{rms})}{\max(n_{S2}^{rms})+\min(n_{S2}^{rms})}$ The root mean square of the photon number is presented by $n_{S2}^{rms}$.}
	\label{vis}
\end{figure}

If the input signal beam is in coherent state $\ket{\alpha e^{i\phi_{\alpha}}}$ and all other modes are in the vacuum state then the expectation value of the output signals becomes: 
\begin{align}
	\label{eq7}
	\ave{S(\delta,\Delta)}=\,\,\,_{IL}\bra{0}_{I0}\bra{0}_{S0}\bra{\alpha e^{i\phi_{\alpha}}}\,\,S(\delta,\Delta)\,\,\ket{\alpha e^{i\phi_{\alpha}}}_{S0}\ket{0}_{I0}\ket{0}_{IL}=\frac{1}{2}\alpha\beta|G|\cos(\phi_{G}(\Delta)+\delta),
\end{align}
where, $\delta=\phi_{\alpha}+\phi_{\beta}$, $\Delta=\phi_{P2}-\phi_{P1}$, $\phi_{G}$ and magnitude $|G| $ is defined as,
\begin{align}
	\label{eq8}
	\phi_{G}(\Delta)=\sin^{-1}(\frac{Tg_1g_2\sin(\Delta-\phi_{T})}{|G|}),
\end{align}
and
\begin{align}
	\label{eq.9}
	|G(\Delta)|^{2}=G_{1}^{2}G_{2}^{2}\left[x^{2}+2x\cos(\Delta-\phi_{T})+1\right],
\end{align}
with $x=Tg_{1}g_{2}(G_{1}G_{2})^{-1}$. Figures \ref{vis}-a and \ref{vis}-b compare the visibility of homodyne detection and conventional Photon Number Counting(PNC).
These figures reveal that the visibility of homodyne detection is a linear function of the object's transmitivity, $T$ and PNC visibility is linear only for low gain regime. The nonlinear response of PNC for high gain regime implies distortions that degrade the quality of the recorded image, This fact is advantageous for roubust performance for homodyne detection across a range of, $T$.

\begin{figure}[tb]
	\centering 
	\subfloat[]{\includegraphics[scale=0.5]{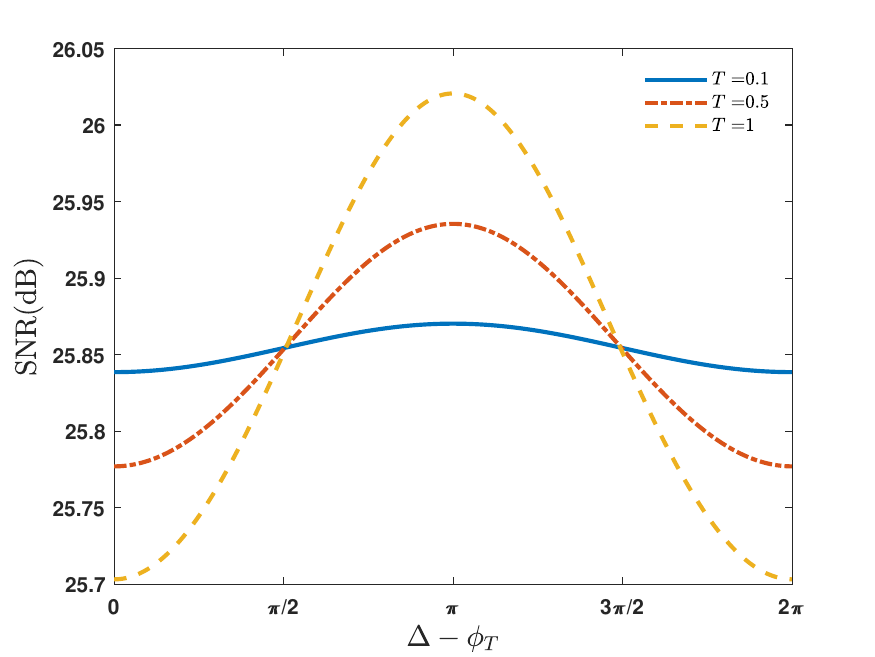}} 
	\subfloat[]{\includegraphics[scale=0.5]{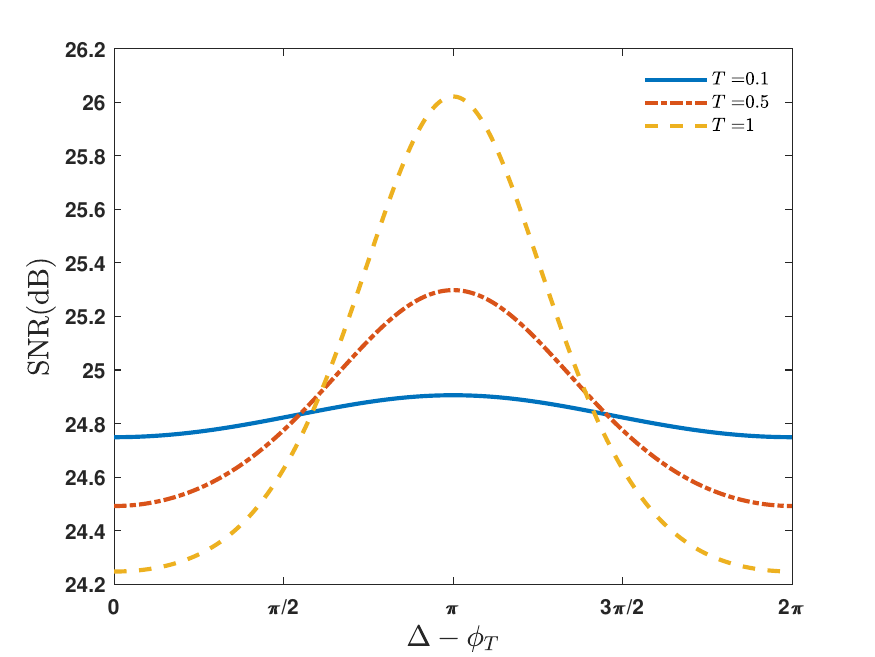}} 
	\hfill
	\subfloat[]{\includegraphics[scale=0.5]{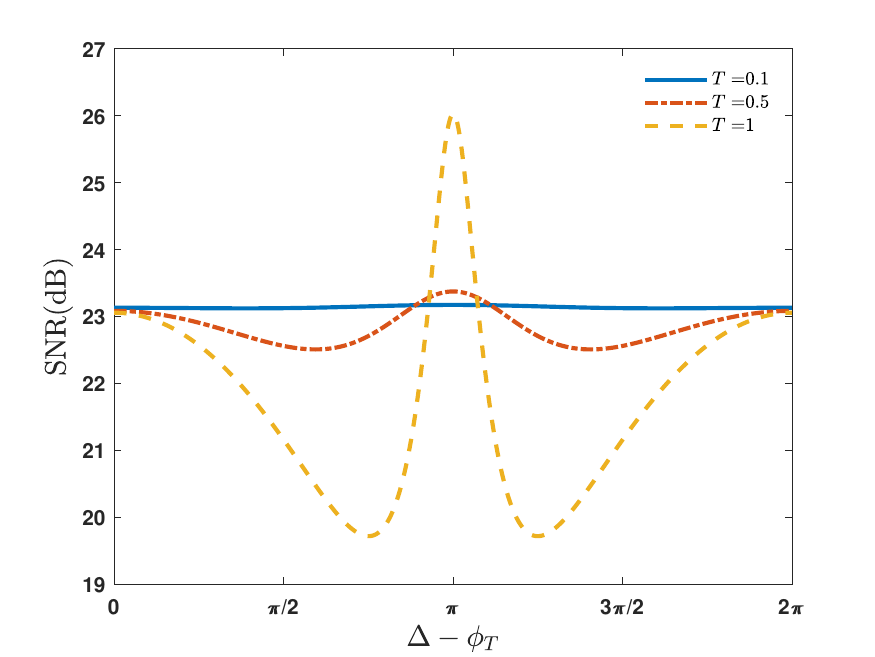}} 
	\subfloat[]{\includegraphics[scale=0.5]{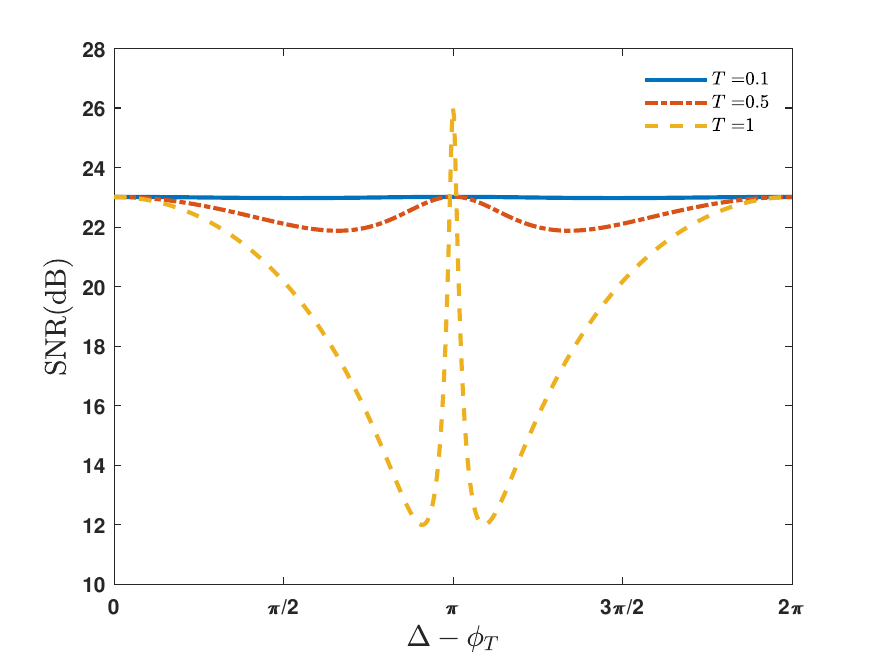}} 
	\caption{homodyne detection SNR as a function of $\Delta-\phi_T$ with $\alpha=100$. (a) $\frac{g_1g_2}{G_1G_2}=0.02$, (b) $\frac{g_1g_2}{G_1G_2}=0.17$, (c) $\frac{g_1g_2}{G_1G_2}=0.75$, (d) $\frac{g_1g_2}{G_1G_2}=0.96$}
	\label{Fig2}
\end{figure}

The signal-to-noise ratio (SNR) is a crucial role in metrology include imaging. SNR is defined as follow:
\begin{align}
	SNR=\frac{\ave{S^2}}{\ave{\Delta S^2}},
\end{align}
here $\ave{S^2}=\frac{\beta^2}{16}\left[4|G|^2\alpha^2\cos^2\left(\phi_{G}+\delta\right)+2|G|^2-1\right]$ is the strength of the signal and $\ave{\Delta S^{2}}=\frac{\beta^{2}}{16}\left(2|G|^{2}-1\right)$ denotes its variance raised by quantum fluctuations.
Interestingly,  $\ave{\Delta S^{2}}$ is related to the optical properties of the object via parameter, G. This fact brings up idea of {\it image construction from noise} in our mind. For the case of  $\alpha\gg1$, both homodyne and PNC detection scheme expression of SNR can be simplified to:
\begin{align}
	\label{eq11}
	\text{SNR}_{HD}=\frac{4\alpha^{2}|G|^{2}}{\left(2|G|^{2}-1\right)}\cos^{2}(\phi_{G}+\delta),
\end{align}
\begin{align}
	\text{SNR}_{PNC}=\frac{\alpha^{2}|G|^{2}}{\left(2|G|^{2}-1\right)}.
\end{align} 
$\text{SNR}_{HD}$ is depicted in \ref{Fig2} as a function of $\Delta-\phi_{T}$. Figure \ref{Fig2} depicts $\text{SNR}_{HD}$ as a function of $\Delta-\phi_{T}$ for the fixed values of the gain. The $\text{SNR}_{HD}$ curve is symmetric around its peak value which occurs at $\Delta-\phi_T=\pi$. Variation of the $\text{SNR}_{HD}$ becomes smoother as the object becomes opaquer.

\begin{figure}[tb]
	\centering 
	\subfloat[]{\includegraphics[scale=0.5]{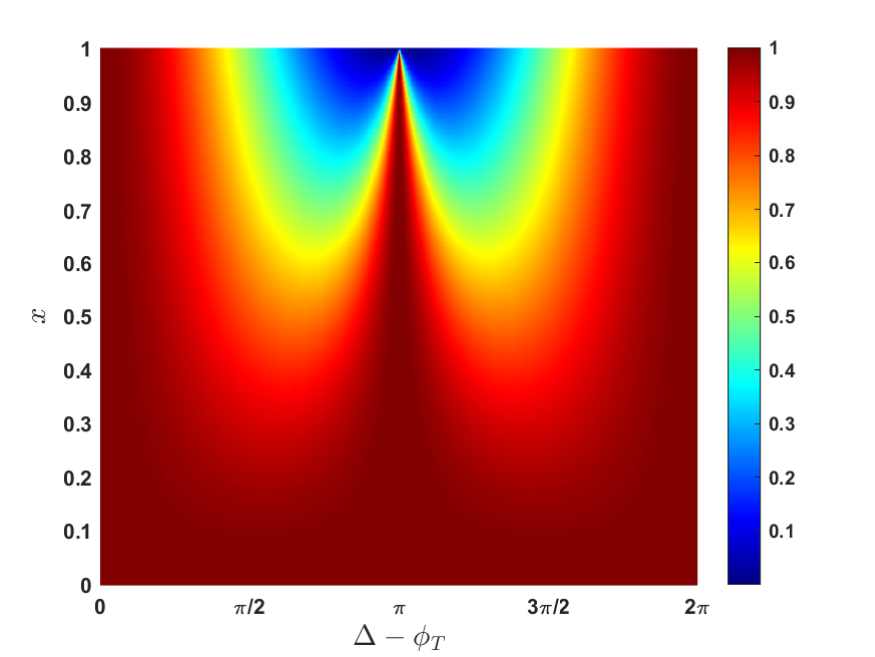}}  
	\subfloat[]{\includegraphics[scale=0.5]{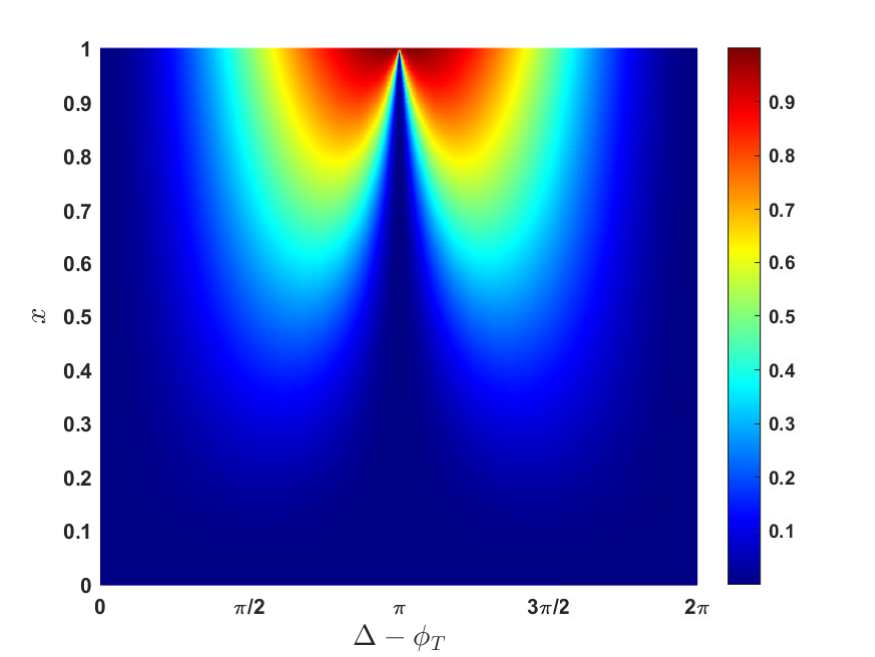}} 
	\caption{Contour plot of $\cos^2(\phi_G+\delta)$ (a)$\delta=0$ and (b) $\delta=\pi/2$.}
	\label{Fig3}
\end{figure}

The ratio $\text{SNR}_{HD}/\text{SNR}_{PC}$ is scaled by a factor of $4\cos^2(\phi_{G}+\delta)$. This quantity is plotted in Fig. \ref{Fig3}. This figure reveals that $\cos^2(\phi_{G} + \delta)$ remains constant for the case of $x \ll 1$. This finding indicates that, $\text{SNR}_{HD}$ can be improved by a factor of order 4 with respect to $SNR_{PNC}$ for transparent objects.

In the following for better illustration, we discuss more about two extremes of  $x\rightarrow0$ (opaque object) and $x\rightarrow1$ (transparent object), separately.

\subsection{Opaque Object, $x\rightarrow 0$}
For transparent object, where $x\rightarrow 0$, $|G|\rightarrow G_{1}G_{2}$ and $\phi_{G}\approx x\sin(\Delta-\phi_{T})$. Retaining \redsout{t}terms up to second order in $x$, the expectation value of $S$ becomes:
\begin{align}
	\label{eq.12}
	\left<S\right>\approx 2\alpha\beta G_{1}G_{2}\left(\cos\delta-\phi_{G}\sin\delta-\frac{1}{2}\phi_{G}^{2}\cos\delta\right).
\end{align}
A small change in $\phi_{T}$ leads to the variation:
\begin{align}
	\label{eq13}
	\left|\delta\left<S\right>\right|=2\alpha\beta G_{1}G_{2}\left|\left(\sin\delta+\phi_{G}\cos\delta\right)\frac{d\phi_{G}}{d\phi_{T}}\right|\delta\phi_{T}.
\end{align}
For $\delta=k\pi/2$, (with integer $k$),.
\begin{align}
	\label{eq.14}
	\left|\delta\left<S\right>\right|=
	\begin{cases}
		2x\alpha\beta G_{1}G_{2}\left|\cos\left(\Delta-\phi_{T}\right)\right|\delta\phi_{T}& k\,\,\text{is odd}\\\\
		x^{2}\alpha\beta G_{1}G_{2}\left|\sin2\left(\Delta-\phi_{T}\right)\right|\delta\phi_{T}& k\,\,\text{is even}
	\end{cases}.
\end{align}
This equation reveals that, in low transmission regime, the small change in the phase leads to a change in output signal which is proportional to  $G_{1}G_{2}$. Therefore sensitivity of a fixed system can be enhanced by nonlinear process gains which is depend on involved parameters such as the phase and amplitude of the pump field. It is also concluded that behavior of the system in this regime is depend on the parity of $k$, such that for the odd values of $k$, output signal scaled by $x$ while for an even value, this quantity is proportional to $x^{2}$. Hence, in this limit we can write:
\begin{align}
	\label{eq15}
	\text{SNR}_{HD}=
	\begin{cases}
		4\alpha^{2}\Gamma\phi_{G}^{2}&k\,\,\text{is odd}\\
		4\alpha^{2}\Gamma&k\,\,\text{is even}
	\end{cases},
\end{align}
where, $\Gamma=\lim_{x\rightarrow 0}(G_{1}G_{2}(2G_{1}G_{2}-1)^{-1})$ , approaching 1 in the low-gain regime and 0.5 in the high-gain regime.  $\text{SNR}_{HD}$ scales as $x^{2}$ for odd $k$ and for even $k$ it is independent of $x$. Eq. \eqref{eq15} implies that $\text{SNR}_{HD}$ can be improved by adjusting the mean photon number in the input signal field of NOPO.
\subsection{High gain, High Transmission, $x\rightarrow 1$}
In the high gain, high transmission regime ($x\rightarrow 1$),we obtain $\phi_{G}=(\Delta-\phi_{T})/2$ and:
\begin{align}
	\label{eq.16}
	|G|=2G_{1}G_{2}\left|\cos\left(\frac{\Delta-\phi_{T}}{2}\right)\right|.
\end{align}
Substituting \eqref{eq.16} into \eqref{eq7} we get the following for the case of $\delta=k\pi/2$.

\begin{align}
	\label{eq.17}
	|\delta\left<S\right>|=
	\begin{cases}
		2\alpha\beta G_{1}G_{2}\left|\cos(\Delta-\phi_{T})\right|\delta\phi_{T}&k\,\, \text{is odd}\\\\
		2\alpha\beta G_{1}G_{2}\left|\sin(\Delta-\phi_{T})\right|\delta\phi_{T}&k\,\, \text{is even}.
	\end{cases}
\end{align}
In this case the sensitivity is independent of $x$ and proportional to the factor $G_{1}G_{2}$. The $\text{SNR}_{HD}$ becomes:
\begin{align}
	\label{eq.18}
	\text{SNR}_{HD}=
	\begin{cases}
		2\alpha^{2}\sin^{2}\left(\frac{\Delta-\phi_{T}}{2}\right)&k\,\,\text{is odd}\\
		2\alpha^{2}\cos^{2}\left(\frac{\Delta-\phi_{T}}{2}\right)&k\,\,\text{is even}\,,
	\end{cases}
\end{align}
which is dependent on adjustable parameter $\alpha$. Hence, $\text{SNR}_{HD}$ can be improved by increasing the mean photon numbers in input signal field of NOPO.
\section{Imaging Protocols }\label{sec.5}
In this section we introduce two data acquisition protocols for image construction. First we present a protocol to extract object information by quantum fluctuation (noise term).
\subsection{Imaging Protocol Using Quantum Fluctuation}\label{QFI}
Since the quantum fluctuation (noise term) depends on the parameter $G$ (see expression for $\ave{\Delta S^2}$), which contains the object information, we can construct the image using the information encoded into quantum fluctuation. To this end we use Eq.\eqref{eq.9}  to rewrite $\ave{\Delta S^2}$ as follows:
\begin{align}
	\label{Neq.1}
	\frac{1}{2G_{1}^{2}G_{2}^{2}}\left[1+\frac{\ave{\Delta S^2(\Delta)}}{\beta^2}\right]=1+x^2+2x\cos(\Delta-\phi_{T}).
\end{align}
It is worthwhile to emphasis that $\Delta$ is experimentally adjustable parameter via precise tuning the phase mismatch of the classical fields pumping the crystals, see figure 1. Therefore the image construction processes can be achieved by the following steps:\\
{\it i)} At first step, the values of $\ave{\Delta S^2 (\Delta)}$ for $\Delta=0,\pi/2,3\pi/2,\pi$ is measured.\\ {\it ii)} The 2nd step is to calculate the image information by the following formulas:
\begin{align}
	\label{Neq.2}
	T=\frac{\sqrt{\left[\ave{\Delta S^{2}(0)}-\ave{\Delta S^{2}(\pi)}\right]^{2}+\left[\ave{\Delta S^{2}(\frac{\pi}{2})}-\ave{\Delta S^{2}(\frac{3\pi}{2})}\right]^{2}}}{4\beta^{2}g_{1}g_{2}G_{1}G_{2}},
\end{align}
and
\begin{align}
	\label{Neq.3}
	\phi_{T}=\tan^{-1}\left(\frac{\ave{\Delta S^{2}(\frac{\pi}{2})}-\ave{\Delta S^{2}(\frac{3\pi}{2})}}{\ave{\Delta S^{2}(0)}-\ave{\Delta S^{2}(\pi)}}\right).
\end{align}
Here we have provided a physical explanation of image formation resulting from noise measurements. In the next section introduce a protocol to extract image information from output signal of homodyne detection.

\subsection{Imaging Protocol using Signal}\label{QSI}
Similarly, the image information ($T$ and $\phi_T$) can be extracted by the output signal.  The signal measured by homodyne detection is governed by Eq. \eqref{eq7} which is proportional to $|G|$. Fortunately, $|G|$ is a function of two experimentally adjustable parameters, $\delta$ and $\Delta$. Hence the following protocol is experimentally executable:\\
{\it i)} Fixing the values $\delta=0,\pi/2$. Under these conditions, the signals measured by detector $\ave{S(\delta, \Delta)}$ can be expressed as:
\begin{subequations}
	\label{eq.19}
	\begin{align}
		\ave{S\left(0,\Delta\right)}=2\alpha\beta|G|\cos\phi_{G},\label{19a}\\
		\ave{S\left(\frac{\pi}{2},\Delta\right)}=-2\alpha\beta|G|\sin\phi_{G}\label{19b}.
	\end{align}
\end{subequations}
Where,
\begin{align}
	\label{eq20}
	\left|G(\Delta)\right|^{2}=\frac{\ave{S(0,\Delta)}^{2}+\ave{S(\frac{\pi}{2},\Delta)}^{2}}{4\alpha^{2}\beta^{2}}.
\end{align}
{\it ii)} Using the fact that $\Delta$ (pumps phase mismatch) is also an adjustable parameter we set $\Delta=0$, $\pi/2$, $\pi$, $3\pi/2$, $\pi$ in Eq. \eqref{eq11} and use results to obtain the information about object,

\begin{align}
	\label{eq21}
	T=\frac{\left(|G(\pi/2)|^{2}-|G(3\pi/2)|^{2}\right)^{2}+\left(|G(0)|^{2}-|G(\pi)|^{2}\right)^{2}}{4G_{1}G_{2}g_{1}g_{2}}.
\end{align}
and
\begin{align}
	\label{eq22}
	\phi_{T}=\tan^{-1}\left[\frac{|G(\pi/2)|^{2}-|G(3\pi/2)|^{2}}{|G(0)|^{2}-|G(\pi)|^{2}}\right],
\end{align}

According to Eq.\eqref{eq20}, $T$ and $\phi_T$ are functions of $\ave{S(0,0)}$, $\ave{S(\pi/2,0)}$, $\ave{S(0,\pi/2)}$, $\ave{S(\pi/2,\pi/2)}$, $\ave{S(0,\pi)}$, $\ave{S(\pi/2,\pi)}$, $\ave{S(0,3\pi/2)}$, $\ave{S(\pi/2,3\pi/2)}$. We emphasize that, although, Eq. \eqref{eq.9} consist of background noises (see the first two term of Eq. \eqref{eq.9}), The last two equations are noise-free. Note that the noise terms in Eq. \eqref{eq22} and the numerator of Eq. \eqref{eq21} are canceled through the calculation, thanks to the homodyne detection. 
\section{Summary and Conclusion}\label{sec.6}
We have presented an imaging setup with undetected photons. The resource photons (signal and idler photon pairs) are generated in two non-linear crystals (posses second order susceptibility, $\chi_1^{(2)}\neq 0$) via SPDC process. In order to enhance the imaging frame rate and reduction of undesired influences of noise (improving SNR), each nonlinear crystal is placed within a confocal cavity, forming a non-degenerate optical parametric oscillatorer (NOPO). The first NOPO is pumped by a coherent beam and seeded with a coherent signal beam. The second NOPO is pumped by a coherent beam (with certain phase relation to the first pump beam) and seeded by signal and idler beams emerging from first NOPO to ensure path identity. The image information can be extracted by a balanced homodyne detection of the signal beams emerged from two NOPOs. Notice that, these signal beams never directly interacts with the object. 
Our extended theoretical treatment, valid beyond the low gain approximation, demonstrates that the SNR improves with increasing mean photon number in the input seed signal. Detailed analysis for the limits $x\rightarrow 0$ (opaque object)and $x\rightarrow 1$(transparent object) reveals that the sensitivity depends on the relative phase of the input seed signal and local oscillator in the former case, while it becomes independent of the parametric gains in the latter. We have also introduced imaging protocols that extract object information from both quantum fluctuation (noise) and the homodyne signal, effectively suppressing background noise.

These results provide a robust theoretical framework for optimizing SNR in quantum imaging with undetected photon (QUIP) and pave the way for noise-free imaging applications in sensitive bio-imaging.

\bibliographystyle{quantum}
\bibliography{myref}

\end{document}